\documentclass[useAMS,usenatbib]{mn2e}
\usepackage{epsf,graphicx,rotating}

\title[The optical spectra of X-shaped radio galaxies]{The optical spectra of X-shaped radio galaxies} 

\author[H. Landt, C. C. Cheung and S. E. Healey]{Hermine Landt$^1$\thanks{E-mail: hlandt@unimelb.edu.au}, Chi C. Cheung$^{2,3}$ and
Stephen E. Healey$^4$ \\ 
$^1$School of Physics, University of Melbourne, Parkville, VIC 3010, Australia \\ 
$^2$NASA Goddard Space Flight Center, Code 661, Greenbelt, MD 20771, USA \\
$^3$Space Science Division, Naval Research Laboratory, Washington, DC 20375, USA \\
$^4$Department of Physics, Stanford University, Stanford, CA 94305, USA }

\begin{document}

\def\la{\mathrel{\hbox{\rlap{\hbox{\lower4pt\hbox{$\sim$}}}\hbox{$<$}}}}
\def\ga{\mathrel{\hbox{\rlap{\hbox{\lower4pt\hbox{$\sim$}}}\hbox{$>$}}}}

\font\sevenrm=cmr7
\def\CaII{Ca~{\sevenrm II}}
\def\Mgb{Mg~{\sevenrm Ib}}
\def\CIII{C~{\sevenrm III}]}
\def\MgII{Mg~{\sevenrm II}}
\def\NeIII{[Ne~{\sevenrm III}]}
\def\NeV{[Ne~{\sevenrm V}]}
\def\OII{[O~{\sevenrm II}]}
\def\OIII{[O~{\sevenrm III}]}
\def\Ha{H{\sevenrm $\alpha$}}
\def\Hb{H{\sevenrm $\beta$}}
\def\Hc{H{\sevenrm $\gamma$}}
\def\Hd{H{\sevenrm $\delta$}}
\def\SII{[S~{\sevenrm II}]}

\newdimen\digitwidth
\setbox0=\hbox{-.}
\digitwidth=\wd0
\catcode `@=\active
\def@{\kern\digitwidth}

\date{Accepted ~~. Received ~~; in original form ~~}

\pagerange{\pageref{firstpage}--\pageref{lastpage}} \pubyear{2010}

\maketitle

\label{firstpage}

\begin{abstract}

  X-shaped radio galaxies are defined by their peculiar large-scale
  radio morphology. In addition to the classical double-lobed
  structure they have a pair of low-luminosity wings that straddles
  the nucleus at almost right angles to the active lobes, thus giving
  the impression of an 'X'. In this paper we study for the first time
  the optical spectral properties of this object class using a large
  sample ($\sim 50$ sources). We find that the X-shaped radio
  population is composed roughly equally of sources with weak and
  strong emission line spectra, which makes them, in combination with
  the well-known fact that they preferentially have radio powers
  intermediate between those of Fanaroff-Riley type I (FR I) and type
  II (FR II) radio galaxies, the archetypal transition population. We
  do not find evidence in support of the proposition that the X-shape
  is the result of a recent merger: X-shaped radio sources do not have
  unusually broad emission lines, their nuclear environments are in
  general not dusty, and their host galaxies do not show signs of
  enhanced star formation. Instead, we observe that the nuclear
  regions of X-shaped radio sources have relatively high
  temperatures. This finding favours models, which propose that the
  X-shape is the result of an overpressured environment.

\end{abstract}


\section{Introduction}

X-shaped radio galaxies are distinguished from the classical
double-lobed radio sources by their peculiar large-scale radio
morphology \citep{Leahy92}. In addition to the primary, active lobes,
defined as such by having an overall higher surface brightness, they
have a secondary pair of wings. These wings are located on each side
of the nucleus and at large angles relative to the primary lobes, thus
giving the impression of an 'X'. A prime example of this object class
is the source 3C~403 \citep[][see their Fig. 13]{Black92}.

Whereas agreement exists about the fact that only the pair of lobes
are actively fed by radio plasma, it is not clear how the wings
formed. The current propositions fall roughly into two main schools of
thought: the wings are either (i) relic emission from previously
active lobes or (ii) backflow emission from the primary, now active
lobes. The first possibility requires a change in the jet direction,
which occured either slowly over time as jet precession
\citep{Ekers78} or suddenly due to a flip of the black hole spin after
a galaxy merger \citep[e.g.,][]{Den02, Mer02, Gop03}. The second
possibility requires the presence of steep pressure gradients in the
environment through which the jet propagates. These gradients are
either due to an asymmetric distribution of the intracluster medium
\citep{Wor95} or to the shape of the host galaxy relative to the jet
direction \citep{Cap02}. A combination of the two possibilities was
proposed early on by \citet{Leahy84}; radio plasma from the hotspots
flows back preferentially into the cavities produced by the radio
lobes of a previously active phase.

Observational evidence has been gathered so far mainly from studies of
the radio spectral index and polarization along the lobes and wings,
and from comparisons between the radio structure and the X-ray and
optical morphologies of the host galaxy and immediate environment. The
radio spectral index is often found to be steeper in the wings than in
the lobes, indicating that the former contain older radio emission
than the latter \citep[e.g.,][]{Murgia01, Rott01, Den02, Lal07}. But
this result is expected in both scenarios, if the wings and lobes are
relic and active emission, respectively, and if the plasma in the
wings has flown first through the lobes. The direction of
polarization, on the other hand, seems to favour the backflow
scenario, since a smooth plasma stream from the active lobe to its
nearest wing is almost always present.

Optical imaging revealed that X-shaped radio sources are mostly
harboured by unperturbed, highly elliptical galaxies and generally
live in poor clusters \citep[e.g.,][]{Ulr96, Cap02, Den02}. This
finding seems to disfavour a scenario, in which a recent merger has
changed the jet direction. In addition, it is always found that the
direction of the active lobes coincides with the {\it major} axis of
the host galaxy, indicating that environmental effects strongly
influence the resultant radio morphology of the backflow \citep{Cap02,
  Sar09}. So far, few X-shaped radio sources have been imaged at X-ray
frequencies, but all such studies favoured the backflow scenario based
on the asymmetries detected in the environment \citep{Wor95, Kraft05,
  Miller09, Hodges10}.

In this paper we present the first systematic analysis of the optical
spectra of X-shaped radio galaxies. To this end we have used a large
sample, which was selected as decribed in Section 2, and for which we
have measured selected spectral quantities as explained in Section
3. With these in hand we have addressed the issue of jet orientation
and general spectral class (Section 4), we have investigated their
broad and narrow emission line regions (Section 5) and we have studied
the properties of the host galaxy (Section 6). Our main results are
summarized in Section 7, where we also present our conclusions.

Throughout this paper we have assumed cosmological parameters $H_0=70$
km s$^{-1}$ Mpc$^{-1}$, $\Omega_{\rm M}=0.3$, and
$\Omega_{\Lambda}=0.7$.

\section{The Sample}

\citet{Cheung07a} selected from the images of the NRAO Very Large
Array (VLA) FIRST survey \citep{Beck95} a sample of $\sim 100$ sources
with clear or tentative X-shaped extended radio morphology. Later,
\citet{Cheung09} slightly extended this sample and defined a subsample
of 50 radio galaxies with bona-fide X-shaped radio morphology and
available optical spectroscopy (see their Table 2). We have considered
here this radio galaxy sample to which we added two sources, namely,
J1430$+$5217, which was identified as a quasar by the Sloan Digital
Sky Survey \citep[SDSS;][]{SloanDR7} but has only narrow emission
lines in its spectrum (see Fig. \ref{sdss}), and J0245$+$1047, whose
X-shaped radio morphology has recently been discovered by
\citet{L08c}. We have also included all X-shaped radio quasars with
optical spectroscopy available to us (7 sources).

\citet{Cheung09} obtained high-quality optical spectroscopy for 27/59
X-shaped radio sources and a further 19/59 sources were
spectroscopically observed by the SDSS (Data Release 5; see
Fig. \ref{sdss}). For another seven sources, namely, J0113$+$0106,
J0115$-$0000, J0245$+$1047, J1309$-$0012, J1606$+$0000, J1952$+$0230,
and J2347$+$0852, the published optical spectra were made available to
us by the authors \citep{Per98, Best99, Tad93, Lacy00}. Table
\ref{general} lists the total spectroscopic sample used in this paper
(53 sources). The six sources excluded are J0009$+$1244, J0058$+$2651,
J1424$+$2637, J1513$+$2607, J1824$+$7420, and J2123$+$2504.

\section{The Spectral Measurements}

We have measured all spectral quantities on rest-frame spectra, which
were corrected for Galactic extinction using $A_V$ values derived from
the Galactic hydrogen column densities of \citet{DL90}. We have
measured the flux and rest-frame equivalent width by integrating the
emission in the line over a local continuum using the IRAF task
\mbox{\sl onedspec.splot} and its option 'e'. In the case of blends
(such as, e.g., \Hc~and \OIII~$\lambda 4363$) we assumed Gaussian
profiles and used the option 'd' of this task to deblend the
individual features. However, this procedure was not necessary in the
case of the SDSS spectra, whose spectral resolution ($R\sim2000$) is
sufficient to resolve all investigated lines.

We have derived $2 \sigma$ upper limits on the rest-frame equivalent
widths and line fluxes of the narrow emission lines \OII~$\lambda
3727$ and \OIII~$\lambda 5007$ when the lines were not detected but
the line position was covered by the spectrum. The non-detection
limits have been calculated assuming a rectangular emission line of
width 1000 km s$^{-1}$.

We have measured the value of the \CaII~break (located at $\sim
4000$~\AA) in spectra $f_{\lambda}$ versus $\lambda$. The \CaII~break
value is defined as $C=1 - (f_-/f_+)$, where $f_-$ and $f_+$ are the
fluxes in the wavelength regions $3750-3950$~\AA~and $4050-4250$~\AA,
respectively. The \CaII~break value is a suitable orientation
indicator for radio-loud active galactic nuclei (AGN): the smaller its
value, the smaller the jet viewing angle, i.e., the angle between the
radio jet and the observer's line of sight \citep[e.g.,][]{L02}.

\section{Orientation and Classification}

\begin{figure}
\centerline{
\includegraphics[scale=0.4]{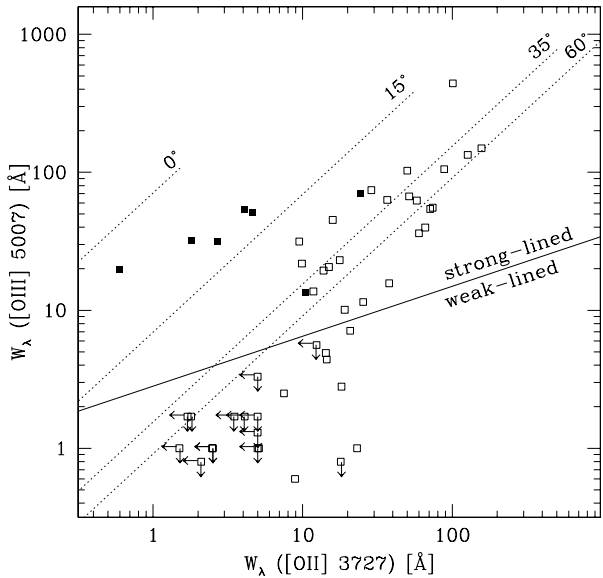}
}
\caption{\label{angles} The rest-frame equivalent widths of
\OIII~$\lambda 5007$ versus \OII~$\lambda 3727$. Filled and open
squares indicate sources with and without broad emission lines,
respectively. Arrows indicate upper limits. The solid line separates
weak-lined (below the line) and strong-lined radio-loud AGN (above the
line). Dotted lines represent loci of constant viewing angles as
labelled.}
\end{figure}

\citet{L04} proposed a physical classification scheme for radio-loud
AGN that takes into account the effects of orientation. This scheme
requires the measurement of only two quantities, namely, the
rest-frame equivalent widths of the narrow emission lines
\OII~$\lambda 3727$ and \OIII~$\lambda 5007$. Based on a bimodality
observed for \OIII, it then separates sources at {\it all} viewing
angles into weak-lined and strong-lined radio-loud AGN.

In Fig. \ref{angles} we show the \OII-\OIII~equivalent width plane for
the X-shaped radio sources in our spectroscopic sample, superposed
with the line dividing the weak-lined and strong-lined classes (solid
line) as well as the loci of constant viewing angles as obtained by
\citet{L04} from their simulations (dotted lines). Our optical spectra
cover the locations of both \OII~and \OIII~for all but one source
(J0702$+$5002) and we list the measurements in Table \ref{general}
(columns (6) and (8)). Two important results become evident from
Fig. \ref{angles}. Firstly, most X-shaped radio sources are viewed at
relatively large angles ($\phi \ga 35^{\circ}$). This result is
supported by their large \CaII~break values. We measure $C\ge0.25$ for
all but six sources and $C\ge0.4$ for roughly half the sample (22/53
sources; see Table \ref{general}, column (4)). The observed result is
intuitive given that projection effects are expected to distort the
X-shape of the diffuse radio emission at small viewing angles, thus
making it difficult to recognize. However, it also means that strongly
relativistically beamed X-shaped radio galaxies (i.e., X-shaped radio
quasars) will be difficult to select based on radio maps.

Secondly, roughly half the sample (23/53 sources) is classified as
weak-lined radio-loud AGN (Table \ref{general}, column (5)), with
about half of these objects (11/23 sources) having no \OII~or
\OIII~emission lines. This result is rather suprising. Since X-shaped
radio sources are generally known to have their active pair of lobes
terminating in pronounced hot spots, as observed for FR II radio
galaxies, one would expect that, like these, they predominantly have
strong optical emission lines. Are then X-shaped radio sources equally
related to FR Is, which are known to have optical spectra with no or
only weak emission lines, or do they simply contain an exceptionally
large number of the otherwise rather rare weak-lined FR IIs
\citep[e.g.,][]{Lai94, Tad98}?

\begin{figure}
\centerline{
\includegraphics[scale=0.4]{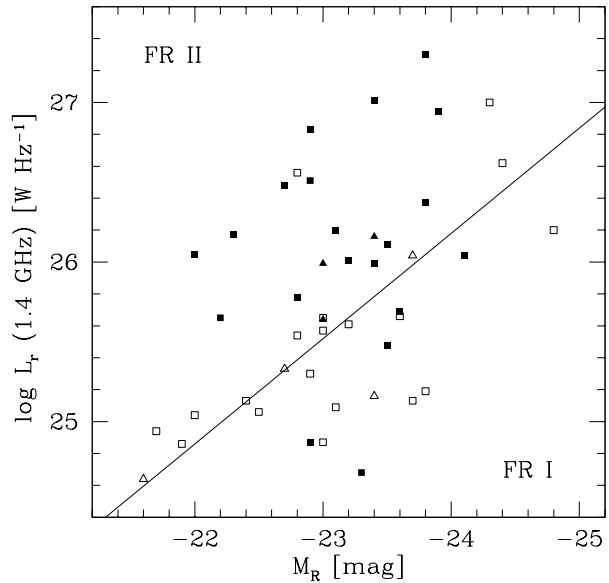}
}
\caption{\label{ledlow} The radio luminosity at 1.4 GHz versus
  absolute $R$ magnitude for X-shaped radio galaxies. Filled and open
  symbols indicate sources classified as strong-lined and weak-lined
  radio-loud AGN, respectively. Squares and triangles indicate sources
  with an FR II and ambiguous radio morphology of the active lobes,
  respectively. The solid line represents the division between FR I
  (below the line) and FR II radio galaxies (above the line) derived
  by \citet{Led96}, scaled to the units of the figure as in
  \citet{Cheung09}.}
\end{figure}

\begin{figure}
\centerline{
\includegraphics[scale=0.4]{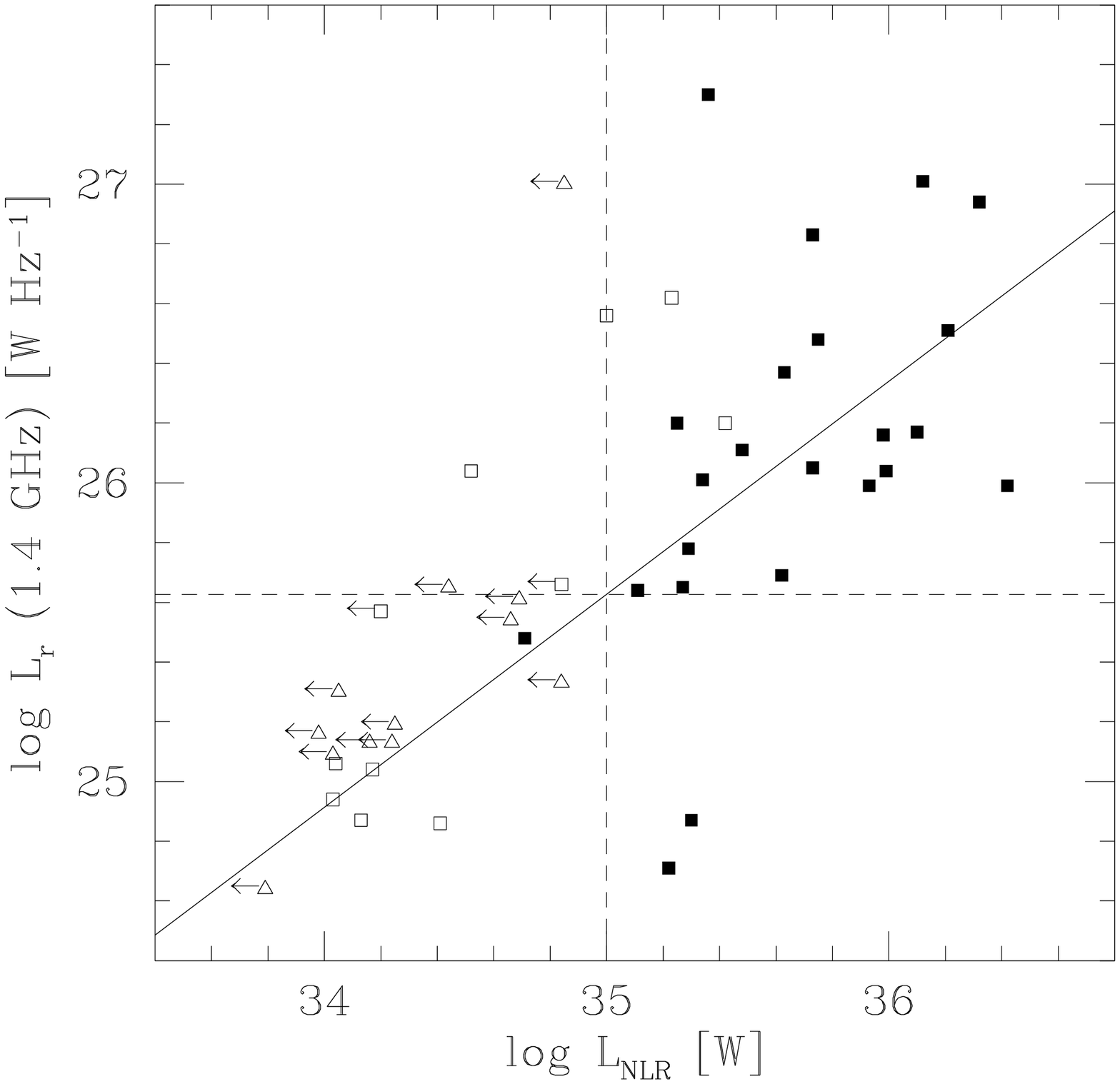}
}
\caption{\label{nlrlext} The radio luminosity at 1.4 GHz versus the
  luminosity of the narrow emission line region for X-shaped radio
  sources viewed at relatively large angles. Symbols are as in
  Fig. \ref{ledlow}. Arrows indicate upper limits. The solid line
  indicates the observed correlation for the strong-lined X-shaped
  radio sources. The dashed lines mark the transition point in type of
  emission line spectrum.}
\end{figure}

In order to answer this question we have plotted in Fig. \ref{ledlow}
the total radio luminosity at 1.4 GHz versus the absolute $R$
magnitude for the strong-lined (filled symbols) and weak-lined
X-shaped radio galaxies (open symbols), which we subdivided based on
the presence or lack of pronounced hotspots in their active lobes into
sources with FR II (squares) and ambiguous (triangles) radio
morphology, respectively. Luminosities were predominantly taken from
\citet{Cheung09} and the radio morphology judged visually from the
FIRST maps and, where available, also from published, deeper radio
maps \citep[see references in][]{Cheung07a}. We have included only
sources viewed at relatively large angles, i.e., without broad
emission lines and with \CaII~break values $C\ge0.25$ (46 sources),
since in their case relativistic beaming effects are expected to be
negligible. Then, Fig. \ref{ledlow} represents the so-called
Ledlow-Owen plot \citep{Led96}, i.e., extended radio emission versus
host galaxy luminosity, in which FR I and FR II galaxies separate
(below and above the solid line, respectively). This separation,
however, does not occur suddenly but rather via a transition region
\citep{Best09}.

Fig. \ref{ledlow} shows that the large majority of weak-lined sources
straddle the FR I/II dividing line, as X-shaped radio galaxies
generally do \citep{Leahy92, Den02, Cheung09}, with only two
(J0148$+$5332 and J1434$+$5906) and five objects (J0001$-$0033,
J0702$+$5002, J0813$+$4347, J1005$+$1154, and J1614$+$2817) clearly in
the FR II and FR I regime, respectively. All but one of these sources
have active lobes with an FR II radio morphology, whereas J0813$+$4347
could be part of the recently identified X-shaped radio population
without pronounced hotspots and similar to FR Is \citep{Sar09}.

It seems then that X-shaped radio galaxies genuinely represent a
transition population, and this in both radio power and emission line
strengths. This transition is illustrated in detail in
Fig. \ref{nlrlext}, where we have plotted the total radio luminosity
at 1.4 GHz versus the luminosity of the narrow emission line region
$L_{\rm NLR}$, the latter calculated from the luminosities of
\OII~$\lambda 3727$ and \OIII~$\lambda 5007$ following
\citet{Raw91}. As Fig. \ref{nlrlext} shows, a clear transition point
between the two classes can be identified at values of $L_{\rm NLR}
\sim 10^{35}$ W and $L_{\rm r} \sim 10^{25.6}$ W Hz$^{-1}$ (dashed
lines). Interestingly, this point lies on the strong ($P=98.6\%$)
linear correlation present for the strong-lined X-shaped radio sources
(solid line). 


\begin{table*}
\caption{\label{general} 
General Properties of the Spectroscopic Sample}
\begin{tabular}{llrcccrrrrl}
\hline
Object Name & Observatory & z & C & class & \multicolumn{2}{c}{\OII~$\lambda 3727$} & 
\multicolumn{2}{c}{\OIII~$\lambda 5007$} & Other Name \\
&&&&& $W_{\lambda}$ & flux & $W_{\lambda}$ & flux \\
&&&&& [\AA] & [erg/s/cm$^2$] & [\AA] & [erg/s/cm$^2$] \\ 
(1) & (2) & (3) & (4) & (5) & (6) & (7) & (8) & (9) & (10) \\
\hline
J0001$-$0033 & SDSS 2.5 m & 0.247 & 0.43 & WL &$<$2.5 &$<$6.68E$-$17 &$<$1.0 &$<$7.51E$-$17 & \\
J0049$+$0059 & SDSS 2.5 m & 0.304 & 0.46 & WL &  18.0 &   2.35E$-$16 &$<$0.8 &$<$4.23E$-$17 & \\
J0113$+$0106 & Shane 3 m  & 0.281 & 0.31 & SL & 101.0 &   2.22E$-$15 & 441.7 &   1.95E$-$14 & \\
J0115$-$0000 & Shane 3 m  & 0.381 & 0.30 & SL & 157.0 &   5.96E$-$16 & 149.3 &   1.16E$-$15 & 4C $-$00.07 \\
J0144$-$0830 & MMT 6.5 m  & 0.181 & 0.39 & WL &$<$1.5 &$<$4.14E$-$17 &$<$1.0 &$<$6.49E$-$17 & \\
J0148$+$5332 & MMT 6.5 m  & 0.290 & 0.25 & WL &$<$5.0 &$<$2.28E$-$16 &$<$1.7 &$<$1.54E$-$16 & 3C 52 \\
J0220$-$0156 & MMT 6.5 m  & 0.173 & 0.35 & SL &  66.0 &   4.03E$-$15 &  39.8 &   6.64E$-$15 & 3C 63 \\
J0245$+$1047 & MMT 4.5 m  & 0.070 & 0.26 & SL &  15.0 &   5.30E$-$15 &  20.6 &   2.05E$-$14 & 4C $+$10.08 \\
J0516$+$2458 & MMT 6.5 m  & 0.063 & 0.33 & SL &  37.9 &   3.55E$-$15 &  15.7 &   4.41E$-$15 & 3C 136.1 \\
J0702$+$5002 & HJST 2.7 m & 0.094 &  ?   & WL &   ?   &    ?         &   5.0 &   8.80E$-$16 & \\
J0805$+$2409 & SDSS 2.5 m & 0.060 & 0.47 & SL &  74.1 &   1.22E$-$14 &  55.4 &   2.40E$-$14 & 3C 192 \\
J0813$+$4347 & SDSS 2.5 m & 0.128 & 0.45 & WL &$<$2.1 &$<$1.65E$-$16 &$<$0.8 &$<$1.61E$-$16 & \\
J0831$+$3219 & SDSS 2.5 m & 0.051 & 0.44 & WL &   7.5 &   1.34E$-$15 &   2.5 &   1.38E$-$15 & 4C $+$32.25 \\
J0845$+$4031 & HET 9.2 m  & 0.429 & 0.31 & SL &   9.9 &   5.25E$-$16 &  21.8 &   2.23E$-$15 & \\
J0859$-$0433 & MMT 6.5 m  & 0.356 & 0.43 & SL &  19.1 &   3.41E$-$16 &  10.1 &   4.59E$-$16 & \\
J0917$+$0523 & HET 9.2 m  & 0.591 & 0.40 & SL &  13.8 &   6.26E$-$16 &  19.4 &   2.01E$-$15 & 4C $+$05.39 \\
J0924$+$4233 & SDSS 2.5 m & 0.227 & 0.42 & WL &$<$4.1 &$<$1.25E$-$16 &$<$1.7 &$<$1.48E$-$16 & \\
J0941$-$0143 & MMT 6.5 m  & 0.384 & 0.45 & WL &  18.2 &   3.16E$-$16 &   2.8 &   1.20E$-$16 & 4C $-$01.19 \\
J0941$+$3944 & SDSS 2.5 m & 0.108 & 0.38 & SL &  15.9 &   1.48E$-$15 &  45.2 &   1.16E$-$14 & 3C 223.1 \\
J1005$+$1154 & SDSS 2.5 m & 0.166 & 0.48 & WL &$<$2.5 &$<$1.60E$-$16 &$<$1.0 &$<$1.99E$-$16 & \\
J1015$+$5944 & SDSS 2.5 m & 0.527 & 0    & SL &   1.8 &   1.86E$-$16 &  32.3 &   2.01E$-$15 & \\
J1018$+$2914 & MMT 6.5 m  & 0.389 & 0.27 & SL &  51.5 &   3.93E$-$16 &  67.0 &   9.87E$-$16 & \\
J1020$+$4831 & SDSS 2.5 m & 0.052 & 0.44 & WL &  23.1 &   2.39E$-$15 &   1.0 &   3.18E$-$16 & 4C $+$48.29 \\
J1101$+$1640 & MMT 6.5 m  & 0.071 & 0.45 & WL &  14.5 &   1.84E$-$15 &   4.4 &   1.44E$-$15 & Abell 1145 \\
J1130$+$0058 & SDSS 2.5 m & 0.133 & 0.26 & SL &  24.2 &   2.59E$-$15 &  69.9 &   1.51E$-$14 & 4C $+$01.30 \\
J1135$-$0737 & MMT 6.5 m  & 0.602 & 0.36 & WL &  20.8 &   1.35E$-$16 &   7.1 &   1.14E$-$16 & \\
J1140$+$1057 & SDSS 2.5 m & 0.081 & 0.41 & WL &   5.1 &   5.72E$-$16 &   1.0 &   3.31E$-$16 & \\
J1206$+$3812 & SDSS 2.5 m & 0.838 & 0    & SL &   2.7 &   2.55E$-$16 &31.5$^{\star}$&2.18E$-$15$^{\star}$& \\
J1207$+$3352 & SDSS 2.5 m & 0.079 & 0.28 & SL &  36.8 &   5.22E$-$15 &  63.1 &   1.85E$-$14 & \\
J1210$-$0341 & MMT 6.5 m  & 0.178 & 0.40 & WL &$<$3.5 &$<$1.12E$-$16 &$<$1.7 &$<$1.38E$-$16 & \\
J1218$+$1955 & MMT 6.5 m  & 0.424 & 0.33 & SL &  50.0 &   2.87E$-$16 & 102.8 &   1.25E$-$15 & 4C $+$20.28 \\
J1228$+$2642 & HET 9.2 m  & 0.201 & 0.47 & WL &$<$5.0 &$<$4.78E$-$16 &$<$1.3 &$<$3.80E$-$16 & \\
J1253$+$3435 & MMT 6.5 m  & 0.358 & 0.30 & SL &   9.5 &   1.04E$-$16 &  31.5 &   6.89E$-$16 & \\
J1309$-$0012 & ESO 3.6 m  & 0.419 & 0.24 & SL &  58.0 &   1.34E$-$15 &  62.4 &   1.89E$-$15 & 4C $+$00.46 \\
J1310$+$5458 & HET 9.2 m  & 0.356 & 0.27 & SL &  71.2 &   1.14E$-$15 &  54.2 &   2.17E$-$15 & \\
J1327$-$0203 & SDSS 2.5 m & 0.183 & 0.48 & WL &   8.9 &   3.49E$-$16 &   0.6 &   7.99E$-$17 & 4C $-$01.29 \\
J1342$+$2547 & HET 9.2 m  & 0.585 & 0.08 & SL &  10.5 &   1.95E$-$16 &  13.4 &   3.03E$-$16 & \\
J1348$+$4411 & MMT 6.5 m  & 0.267 & 0.28 & WL &$<$12.4&$<$1.61E$-$16 &$<$5.6 &$<$1.43E$-$16 & \\
J1357$+$4807 & HET 9.2 m  & 0.383 & 0.26 & SL &  88.3 &   1.27E$-$15 & 105.5 &   3.00E$-$15 & \\
J1406$-$0154 & HET 9.2 m  & 0.641 & 0.40 & SL &  11.8 &   6.03E$-$17 &  13.7 &   1.71E$-$16 & 4C $-$01.31 \\
J1406$+$0657 & HET 9.2 m  & 0.550 & 0    & SL &   0.6 &   3.36E$-$16 &19.7$^{\dagger}$&6.89E$-$15& \\
J1430$+$5217 & SDSS 2.5 m & 0.367 & 0.28 & SL & 127.0 &   2.02E$-$15 & 133.8 &   3.65E$-$15 & \\
J1434$+$5906 & HET 9.2 m  & 0.538 & 0.45 & WL &  14.3 &   7.04E$-$17 &   4.9 &   5.01E$-$17 & \\
J1444$+$4147 & SDSS 2.5 m & 0.188 & 0.44 & WL &   1.8 &   7.64E$-$17 &$<$1.7 &$<$2.02E$-$16 & \\
J1456$+$2542 & HET 9.2 m  & 0.536 & 0.39 & WL &$<$1.7 &$<$2.19E$-$17 &$<$1.7 &$<$5.18E$-$17 & \\
J1600$+$2058 & MMT 6.5 m  & 0.174 & 0.37 & SL &  25.4 &   1.08E$-$15 &  11.5 &   1.25E$-$15 & \\
J1606$+$0000 & ESO 3.6 m  & 0.059 & 0.41 & WL &$<$5.0 &$<$1.18E$-$15 &$<$1.0 &$<$6.59E$-$16 & 4C $+$00.58 \\
J1606$+$4517 & HET 9.2 m  & 0.556 & 0.37 & SL &  59.9 &   5.15E$-$16 &  36.2 &   6.72E$-$16 & \\
J1614$+$2817 & MMT 6.5 m  & 0.108 & 0.42 & WL &$<$5.0 &$<$2.25E$-$16 &$<$3.3 &$<$3.63E$-$16 & \\
J1625$+$2705 & SDSS 2.5 m & 0.526 & 0    & SL &   4.1 &   3.97E$-$16 &  53.9 &   3.01E$-$15 & \\
J1952$+$0230 & ESO 2.2 m  & 0.059 & 0.40 & SL &  28.8 &   1.36E$-$14 &  74.3 &   8.56E$-$14 & 3C 403 \\
J2157$+$0037 & SDSS 2.5 m & 0.391 & 0.35 & SL &  17.7 &   2.56E$-$16 &  23.1 &   7.47E$-$16 & \\
J2347$+$0852 & KPNO 2.1 m & 0.292 & 0    & SL &   4.6 &   2.20E$-$15 &  51.1 &   2.37E$-$14 & \\     
\hline		 
\end{tabular}

\medskip

\parbox[]{15.5cm}{The columns are: (1) object name; (2) observatory at
  which the spectrum was obtained; (3) redshift; (4) \CaII~break value
  (measured in spectra $f_{\lambda}$ versus $\lambda$); (5)
  classification following \citet{L04}, where WL: weak-lined radio-loud
  AGN, SL: strong-lined radio-loud AGN; (6) rest-frame equivalent width
  and (7) flux of \OII~$\lambda 3727$; (8) rest-frame equivalent width
  and (9) flux of \OIII~$\lambda 5007$; and (10) common object name.}

\medskip

\parbox[]{15.5cm}{$\star$ derived from \OIII~$\lambda 4959$; 
$\dagger$ rest-frame equivalent width relative to general continuum}

\end{table*}

\section{The Emission Line Regions}

Seven sources in our spectroscopic sample have broad emmission lines
and we discuss their properties below (Section \ref{belr}). For these
(with the exception of J1406$+$0657) and a further 22 sources we
detect besides \OII~$\lambda 3727$ and \OIII~$\lambda 5007$ also other
useful narrow emission lines, which we analyze in Section
\ref{nelr}. We do not further discuss four sources (J1018$+$2914,
J1253$+$3435, J1309$-$0012, and J1406$-$0154) that have only H$\beta$
$\lambda 4861$, \OII~and \OIII. The remaining 20 sources in our sample
have either no emission lines or none besides \OII~and \OIII.

\subsection{The Broad Emission Lines} \label{belr}


\begin{table}
\caption{\label{fwhm} Broad Emission Line Widths}
\begin{tabular}{lccc}
\hline
Object Name & \Ha~$\lambda 6563$ & \Hb~$\lambda 4861$ & \MgII~$\lambda 2798$ \\
& FWHM & FWHM & FWHM \\
& [km/s] & [km/s] & [km/s] \\ 
(1) & (2) & (3) & (4) \\
\hline
J1015$+$5944 &  ?   &  5727 & 3656 \\
J1130$+$0058 & 7241 &   --  &  ?   \\
J1206$+$3812 &  ?   &  8048 & 7473 \\
J1342$+$2547 &  ?   &   --  & 9864 \\
J1406$+$0657 &  ?   & 12053 & 8792 \\
J1625$+$2705 &  ?   &  7079 & 4074 \\
J2347$+$0852 & 8420 &  7926 &  ?   \\ 
\hline
\end{tabular}
\end{table}

Many popular explanations for the genesis of X-shaped radio sources
require a galaxy merger, which is expected to yield a binary
supermassive black hole. Such a binary can have two broad-line systems
and/or two narrow-line systems, if both supermassive black holes are
quasars and their spatial separation is large enough
\citep[e.g.,][]{Pet87, Gas96, Bor09, Shen09}. In particular the two
broad-line systems are expected to be observable as a single but
unusually broad emission line.

In order to assess if X-shaped radio quasars have unusually broad
emission lines, we have measured the full width at half-maximum (FWHM)
of the most prominent lines present in our spectra, namely, the Balmer
lines H$\alpha$ $\lambda 6563$ and H$\beta$ $\lambda 4861$, and
\MgII~$\lambda 2798$. The narrow emission line components of H$\alpha$
and H$\beta$ are pronounced in all objects and were removed prior to
measuring the broad component. Such a correction was not
straightforward for \MgII~and we left the line unchanged. In Table
\ref{fwhm} we list the measurements, where a question mark indicates
that the position of the line was not covered by the spectrum. For
J1130$+$0058 we observe broad H$\alpha$, but only narrow H$\beta$. The
lack of a broad H$\beta$ component in this source is most likely due
to dust obscuration (see Section \ref{nelr} and \citet{Wang03,
  Zhang07}). For J1342$+$2547 a broad H$\beta$ line appears to be
present, but the spectrum is too noisy to reliably measure its width.

The observed broad emission line widths of X-shaped radio quasars span
a range similar to that generally found for lobe-dominated radio
quasars \citep{Broth96, Corb97, Aars05}. In this respect we note that
the relatively large H$\beta$ width of J1406$+$0657 could be due to a
strong ``red shelf'', an emission feature known to be asscociated with
broad H$\beta$ in some quasars \citep[e.g.,][]{DeR85, Marz96}. This
result argues against merger models, in which the two black holes are
still separated. However, it does not exclude any binary model, if it
is assumed that one of the black holes has no emission line system.

\subsection{The Narrow Emission Lines} \label{nelr}

The narrow emission lines probe the nuclear gas environment on scales
up to a few kpc and can reveal important information that might
constrain models for the origin of the X-shaped radio morphology. For
example, a perturbed and rather dusty environment is expected in the
case of a recent galaxy merger. On the other hand, the large pressure
gradients required by the backflow model could have also increased the
gas densities and/or temperatures.

In the following we estimate the nuclear dust extinction using the
three strongest hydrogen lines (Section \ref{dust}) and derive gas
electron densities and temperatures based on sulphur and oxygen
emission line ratios, respectively (Section \ref{denstemp}). Table
\ref{narrow} lists the relevant flux measurements (28/53 sources).

\subsubsection{Dust Extinction} \label{dust}

\begin{figure}
\centerline{
\includegraphics[scale=0.4]{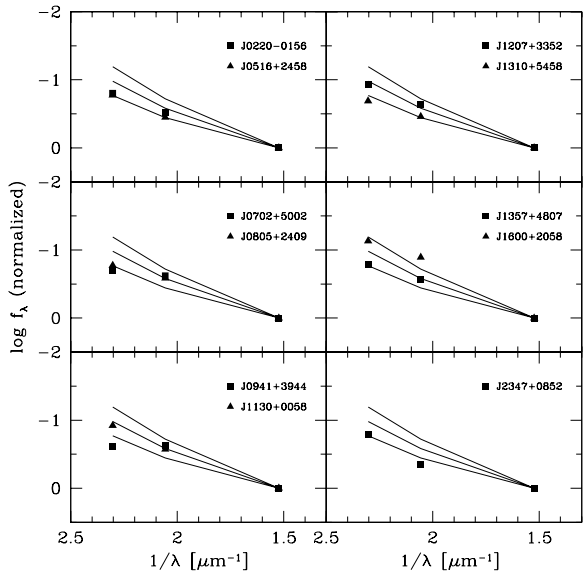}
}
\caption{\label{caseB} Balmer emission line fluxes compared to
expectations from Case B recombination (for a temperature of
$T=15,000$ K and an electron density of $n_e=10^4$ cm$^{-3}$). From
bottom to top, the dust extinction assumed was $A_V=0$, 1, and 2
mag. The measurement points and the Case B solid lines were normalized
to the flux of H$\alpha$.}
\end{figure}

We detect all three Balmer emission lines H$\alpha$, H$\beta$ and
H$\gamma$ in 11/28 sources. These sources are all but one
(J0702$+$5002) classified as strong-lined radio-loud AGN. In
Fig. \ref{caseB} we compare their measured line fluxes with the
predictions from Case B recombination without dust (lower solid line),
and including a dust extinction of $A_V = 1$ and 2 mag (middle and
upper solid lines, respectively). The values expected from Case B
recombination were calculated using the Cloudy photoionization
simulation code \citep[last described by][]{Cloudy} and assuming a
temperature of $T=15,000$ K and an electron density of $n_e=10^4$
cm$^{-3}$. The $A_V$ values were transformed into $A_{\lambda}$ values
using the analytical expression for the interstellar extinction curve
of \citet{Car89} and assuming a parameter $R_V=3.1$. Fig. \ref{caseB}
shows the emission line fluxes and the Case B lines normalized such
that the H$\alpha$ flux is unity.

In all but three sources we estimate the dust extinction to be
negligible. In two sources (J1130$+$0058 and J1207$+$3352) we obtain a
dust extinction of the order of $A_V \sim 1$ mag and the highest
extinction ($A_V \sim 2$ mag) is observed for J1600$+$2058. The amount
of dust observed in the X-shaped radio quasar J1130$+$0058 may explain
why this source displays a broad H$\alpha$ but only a narrow H$\beta$
emission line.

The general lack of large amounts of dust in the nuclear environments
of X-shaped radio galaxies indicates that any galaxy merger must have
happened a long time ago. This conclusion is similar to that drawn by
authors who studied the host galaxies of X-shaped sources with optical
imaging. They find them to be almost perfectly elliptical, without the
usual signs of a recent interaction, such as, e.g., tidal tails or
extended dust lanes \citep[e.g.,][]{Wirth82, Breu83, Cap02, Cheung07b,
  Sar09}.

\subsubsection{Electron Densities and Temperatures} \label{denstemp}

\begin{figure}
\centerline{
\includegraphics[scale=0.4]{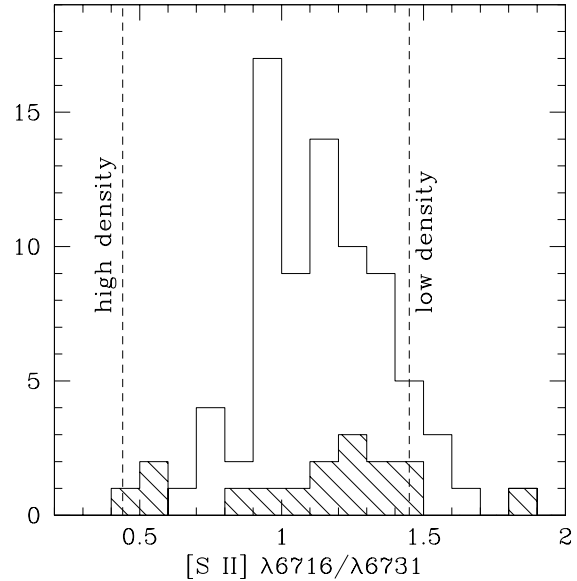}
}
\caption{\label{dens} Histogram of the flux ratio between the sulphur
  lines \SII~$\lambda \lambda 6716, 6731$ for X-shaped sources
  (shaded) and radio galaxies from the 3CR survey. This ratio is
  sensitive to the electron density in the intermediate regime of
  $n_e\approx10-10^5$ cm$^{-3}$, marked by the dashed lines.}
\end{figure}

\begin{figure}
\centerline{
\includegraphics[scale=0.4]{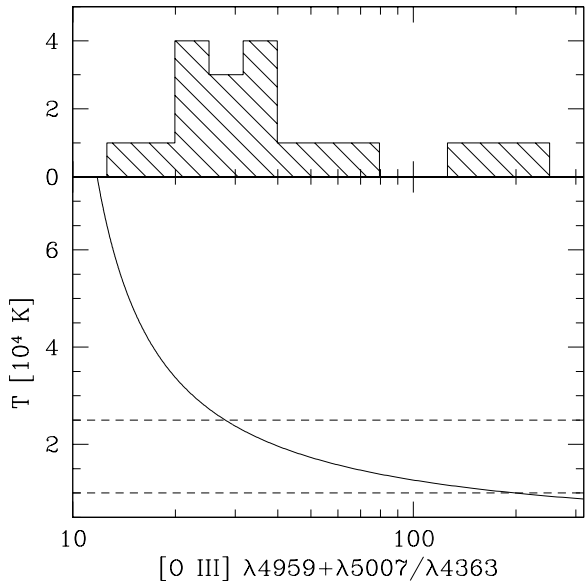}
}
\caption{\label{temp} Top panel: Histogram of the flux ratio between
  the oxygen doublet \OIII~$\lambda \lambda 4959, 5007$ and
  \OIII~$\lambda 4363$ for X-shaped sources. Bottom panel: Theoretical
  relation between this ratio and the electron temperature calculated
  following \citet{Osterbrock2} and assuming an electron density of
  $n_e=10^4$ cm$^{-3}$. The dashed lines mark the temperature range
  usually derived for AGN narrow emission line regions.}
\end{figure}

The flux ratio between the lines of the sulphur doublet \SII~$\lambda
\lambda 6716, 6731$ is a suitable indicator of electron density (the
higher its value, the lower the density) in the intermediate regime of
$n_e\approx10-10^5$ cm$^{-3}$ \citep[e.g.,][]{Peterson}. We detect the
two sulphur lines in 16/28 sources, of which equal numbers are
classified as weak-lined and strong-lined AGN. In Fig. \ref{dens} we
compare their \SII~line ratios with those observed for a sample of 76
'normal' radio galaxies from the 3CR survey \citep{But09}. We note
that we have excluded from the 3CR comparison sample sources in common
with our sample. Fig. \ref{dens} shows that X-shaped sources reach
both higher and lower electron densities than is typical of radio
galaxies, however, according to a Kolmogorov-Smirnov (KS) test the
distributions of the two groups are not significantly different
($P<95\%$).

The flux ratio between the sum of the oxygen doublet \OIII~$\lambda
\lambda 4959, 5007$ and the oxygen line \OIII~$\lambda 4363$ is a
suitable indicator of electron temperature (the higher its value, the
lower the temperature). We detect the \OIII~$\lambda 4363$ line in
19/28 sources. These sources are all classified as strong-lined
radio-loud AGN. In Fig. \ref{temp} (top panel) we have plotted the
distribution of the observed oxygen ratios, which suggests that the
large majority of X-shaped sources have relatively high electron
temperatures. Using a theoretical relation calculated following
\citet{Osterbrock2} and assuming an electron density of $n_e=10^4$
cm$^{-3}$ (Fig. \ref{temp}, lower panel), we estimate for 13/19
sources temperatures well in excess of the typical value of $T \sim
15,000$ K, with eight sources even exceeding the upper boundary of $T
\sim 25,000$ K usually derived for AGN narrow emission line regions
\citep[e.g.,][]{Peterson}.

We note that the \OIII~$\lambda 4363$ emission line is usually
difficult to measure in AGN, since it is often weak and can blend with
$H\gamma$. However, in particular in the aforementioned 13 sources, we
observe this line to be relatively strong with typically a rest-frame
equivalent width of $W_{\lambda} \sim 5$~\AA~for a signal-to-noise
ratio in the continuum of $S/N \sim 20$. Furthermore, the resolution
of the majority of the spectra is sufficient to reliably separate it
from $H\gamma$.

Our results hint at increased pressures being present in the nuclear
environments of X-shaped radio galaxies, mainly caused by elevated
temperatures. Such enhanced pressures are necessary in the backflow
model of \citet{Cap02}, which proposed that wings emerge in addition
to lobes when the jet is orientated along the major axis of the host
galaxy. Then, in such a case, the jet lobe becomes overpressured and
radio plasma will eventually escape along the galaxy minor axis. In
this respect we note that, contrary to our results, the X-ray studies
of \citet{Hodges10} found no significant differences in temperature
(or density) between a sample of eight X-shaped sources and a sample
of 18 'normal' radio galaxies.


\begin{table*}
\caption{\label{narrow} 
Narrow Emission Line Fluxes}
\begin{tabular}{llcccccccc}
\hline
Object Name & Observatory & class & \Ha~$\lambda 6563$ & \Hb~$\lambda 4861$ & \Hc~$\lambda 4340$ & 
\SII~$\lambda 6716$ & \SII~$\lambda 6731$ & \OIII~$\lambda 4363$ & \OIII \\
&&& flux & flux & flux & flux & flux & flux & flux \\
&&& [erg/s/cm$^2$] & [erg/s/cm$^2$] & [erg/s/cm$^2$] & [erg/s/cm$^2$] & [erg/s/cm$^2$] & 
[erg/s/cm$^2$] & ratio \\ 
(1) & (2) & (3) & (4) & (5) & (6) & (7) & (8) & (9) & (10) \\
\hline
J0049$+$0059 & SDSS 2.5 m & WL & 7.68E$-$17 & --         & --         & 8.14E$-$17 & 6.58E$-$17 & --         & --  \\
J0113$+$0106 & Shane 3 m  & SL &  ?         & 1.47E$-$15 & 7.05E$-$16 &  ?         &  ?         & 1.96E$-$16 & 133 \\
J0115$-$0000 & Shane 3 m  & SL &  ?         & 2.18E$-$16 & 1.40E$-$16 &  ?         &  ?         & 4.56E$-$17 &  34 \\
J0220$-$0156 & MMT 6.5 m  & SL & 3.75E$-$15 & 1.15E$-$15 & 5.91E$-$16 & 6.03E$-$16 & 6.35E$-$16 & 5.63E$-$16 &  16 \\
J0245$+$1047 & MMT 4.5 m  & SL & 4.60E$-$15 & --         & 6.90E$-$16 & --         & --         & 1.60E$-$15 &  17 \\
J0516$+$2458 & MMT 6.5 m  & SL & 6.07E$-$15 & 2.16E$-$15 & 1.02E$-$15 & 3.37E$-$15 & 3.28E$-$15 & --         & --  \\
J0702$+$5002 & HJST 2.7 m & WL & 1.53E$-$15 & 3.66E$-$16 & 3.10E$-$16 & 7.11E$-$16 & 1.22E$-$15 & --         & --  \\
J0805$+$2409 & SDSS 2.5 m & SL & 9.23E$-$15 & 2.40E$-$15 & 1.55E$-$15 & 3.53E$-$15 & 2.69E$-$15 & 1.11E$-$15 &  29 \\
J0831$+$3219 & SDSS 2.5 m & WL & 1.00E$-$15 & --         & --         & 1.09E$-$15 & 7.36E$-$16 & --         & --  \\
J0941$+$3944 & SDSS 2.5 m & SL & 3.81E$-$15 & 9.04E$-$16 & 9.23E$-$16 & 1.02E$-$15 & 8.38E$-$16 & 6.02E$-$16 &  26 \\
J1015$+$5944 & SDSS 2.5 m & SL &  ?         & 2.54E$-$16 & 1.72E$-$16 &  ?         &  ?         & 1.27E$-$16 &  21 \\
J1020$+$4831 & SDSS 2.5 m & WL & 1.29E$-$15 & --         & --         & 9.44E$-$16 & 7.84E$-$16 & --         & --  \\
J1101$+$1640 & MMT 6.5 m  & WL & 9.06E$-$16 & --         & --         & 9.86E$-$16 & 7.52E$-$16 & --         & --  \\
J1130$+$0058 & SDSS 2.5 m & SL & 5.66E$-$15 & 1.51E$-$15 & 6.77E$-$16 & 1.13E$-$15 & 9.70E$-$16 & 3.08E$-$16 &  65 \\
J1140$+$1057 & SDSS 2.5 m & WL & 3.20E$-$16 & --         & --         & 2.32E$-$16 & 4.36E$-$16 & --         & --  \\
J1206$+$3812 & SDSS 2.5 m & SL &  ?         & 1.79E$-$16 & 9.98E$-$17 &  ?         &  ?         & 8.19E$-$17 &  35 \\
J1207$+$3352 & SDSS 2.5 m & SL & 8.16E$-$15 & 1.90E$-$15 & 9.51E$-$16 & 2.63E$-$15 & 2.24E$-$15 & 6.38E$-$16 &  39 \\
J1218$+$1955 & MMT 6.5 m  & SL &  ?         & 1.25E$-$16 & 5.98E$-$17 &  ?         &  ?         & 7.57E$-$17 &  22 \\
J1310$+$5458 & HET 9.2 m  & SL & 6.08E$-$16 & 2.11E$-$16 & 1.24E$-$16 & --         & --         & 1.19E$-$16 &  24 \\
J1327$-$0203 & SDSS 2.5 m & WL & 1.34E$-$16 & --         & --         & 2.54E$-$16 & 1.41E$-$16 & --         & --  \\
J1342$+$2547 & HET 9.2 m  & SL &  ?         & 6.30E$-$17 & 5.90E$-$17 &  ?         &  ?         & 1.91E$-$17 &  21 \\
J1357$+$4807 & HET 9.2 m  & SL & 1.41E$-$15 & 3.78E$-$16 & 2.27E$-$16 & --         & --         & 2.36E$-$17 & 169 \\
J1430$+$5217 & SDSS 2.5 m & SL &  ?         & 5.22E$-$16 & 3.17E$-$16 &  ?         &  ?         & 1.37E$-$16 &  36 \\
J1600$+$2058 & MMT 6.5 m  & SL & 1.19E$-$15 & 1.53E$-$16 & 8.77E$-$17 & 6.18E$-$16 & 4.22E$-$16 & 3.35E$-$17 &  50 \\
J1606$+$0000 & ESO 3.6 m  & WL & 2.08E$-$16 & --         & --         & 8.00E$-$16 & 9.63E$-$16 & --         & --  \\
J1625$+$2705 & SDSS 2.5 m & SL & ?          & 3.66E$-$16 & 1.83E$-$16 &  ?         &  ?         & 1.47E$-$16 &  27 \\
J1952$+$0230 & ESO 2.2 m  & SL & ?          & 5.57E$-$15 & 4.43E$-$15 &  ?         &  ?         & 2.04E$-$15 &  56 \\
J2347$+$0852 & KPNO 2.1 m & SL & 2.46E$-$15 & 1.11E$-$15 & 4.02E$-$16 & 2.12E$-$16 & 4.74E$-$16 & 1.28E$-$16 & 247 \\
\hline
\end{tabular}

\medskip

\parbox[]{18cm}{The columns are: (1) object name; (2) observatory at which the spectrum was obtained; (3) 
  classification; (4) - (9) fluxes of the narrow emission
  lines as labeled, where ?: line position not covered by spectrum, and
  -- : line position covered by spectrum, but line not detected; and (10) flux ratio \OIII~$\lambda 4959+\lambda    5007$/$\lambda 4363$, calculated assuming the theoretical flux ratio \OIII~$\lambda 5007$/$\lambda 4959 \approx 3$.}

\end{table*}

\section{The Host Galaxy} \label{host}

\begin{figure}
\centerline{
\includegraphics[scale=0.4]{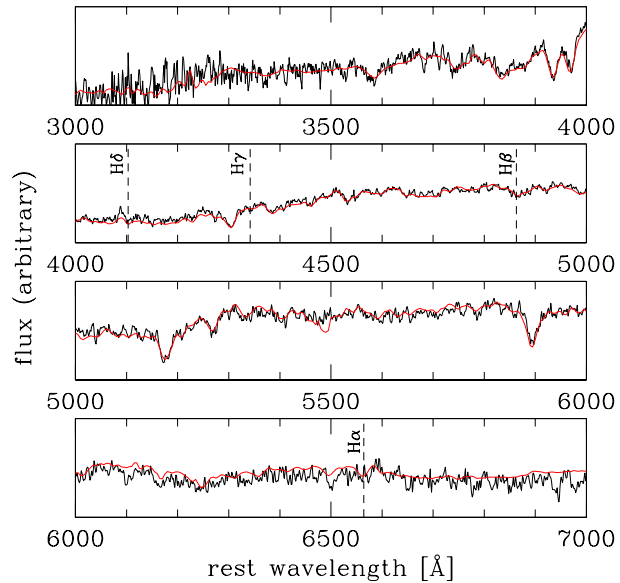}
}
\caption{\label{composite} Composite host galaxy spectrum for X-shaped
  radio sources without \OII~or \OIII~emission lines and \CaII~break
  values of $C\ge0.4$ (black solid line), compared to the elliptical
  galaxy template of \citet{Man01} (red solid line). The dashed lines
  mark the locations of the strongest Balmer hydrogen absorption
  lines.}
\end{figure}

The spectrum of the host galaxy can also put constraints on models for
the origin of the X-shaped radio morphology. In particular, if a
recent galaxy merger was the source of a change in jet direction, we
expect to see spectral signs of starburst activity, such as strong
hydrogen absorption lines that are typical of young stars or enhanced
continuum flux at short wavelengths.

In order to investigate the properties of the host galaxy spectra of
X-shaped radio sources, we have created a composite. For this purpose
we have used only those sources in our sample that had a 'pure' host
galaxy spectrum, i.e., had no \OII~or \OIII~emission lines and a
\CaII~break value of $C\ge0.4$ (8 objects). The latter constraint
ensured that the blue part of the spectrum was not enhanced by the jet
at any perceptible level \citep[see Fig. 1 in][]{L02}, which could
otherwise be wrongly interpreted as a sign of enhanced star formation.

Fig. \ref{composite} shows the resulting composite host galaxy
spectrum (black solid line), compared to the elliptical galaxy
template of \citet{Man01} (red solid line). There are no pronounced
differences present between the two spectra, and in particular none of
the Balmer hydrogen absorption lines H $\alpha$, H $\beta$, H $\gamma$
or H $\delta$ appear unusually strong.

This result indicates, similar to that of Section \ref{dust}, that any
galaxy merger must have happened a considerable time ago. Since the
starburst should be observable for at least a few million years after
its initiation, which is of the order of the electron radiative
lifetimes, this then means that the secondary wings are unlikely to be
plasma from the old jets and are rather backflow from the currently
active lobes.

\section{Summary and Conclusions}

We have analysed the optical spectral properties of a large sample (53
objects) of X-shaped radio sources. The large majority of our sample
are radio galaxies, but seven sources are radio quasars (i.e., have
broad emission lines). We have considered the general relation between
radio and emission line power, searched for signs of recent merger
activity and looked for probes of the nuclear environment. Our main
results can be summarized as follows:

\vspace*{0.1cm}

(i) The X-shaped radio population contains roughly equal numbers of
sources with a weak and a strong emission line spectrum. Given also
that their radio powers mostly straddle the dividing line between FR I
and FR II radio galaxies, we show that this kind of sources are the
archetypal transition population. We observe a clear transition point
in both narrow emission line and radio power at $L_{\rm NLR} \sim
10^{35}$ W and $L_{\rm r} \sim 10^{25.6}$ W Hz$^{-1}$, respectively.

(ii) We have searched for signs of a recent galaxy merger, such as,
unusually broad emission lines that would indicate a
(large-separation) binary black hole, dusty nuclear environments as
probed by the narrow emission lines, and signs of enhanced star
formation in the host galaxy. All three approaches gave negative
results, suggesting that any merger must have occurred a considerable
time ago (longer than the electron radiative lifetimes), thus making
it improbable that the pair of wings are relic radio emission.

(iii) We have probed the electron densities and temperatures in the
nuclear environments of X-shaped radio sources using narrow emission
line diagnostics. We find that the majority of sources have relatively
high temperatures ($T\ga15,000$ K). This supports the scenario where
overpressured environments rather than recent mergers seem to be the
cause for the X-shape radio morphology.

\section*{Acknowledgments}

We thank Mark Lacy, Philip Best and Clive Tadhunter for providing
spectra in electronic format. We are indebted to Kirk Korista for
providing the results from the Cloudy photoionization code
calculations. An anonymous reviewer provided useful suggestions that
helped us to improve the paper.

\bibliography{references}

\begin{thebibliography}{}

\bibitem[\protect\citeauthoryear{Aars, Hough, Yu, Linick, Beyer, Vermeulen \&
  Readhead}{Aars et~al.}{2005}]{Aars05}
Aars C.~E.,  Hough D.~H.,  Yu L.~H.,  Linick J.~P.,  Beyer P.~J.,  Vermeulen
  R.~C.,    Readhead A. C.~S.,  2005, AJ, 130, 23

\bibitem[\protect\citeauthoryear{Abazajian, Adelman-McCarthy, Ag\"ueros, Allam,
  Allende~Prieto, An, Anderson, Anderson et~al.,}{Abazajian
  et~al.}{2009}]{SloanDR7}
Abazajian K.~N.,  Adelman-McCarthy J.~K.,  Ag\"ueros M.~A.,  Allam S.~S.,
  Allende~Prieto C.,  An D.,  Anderson K. S.~J.,  Anderson S.~F.,    et~al.,
  2009, ApJS, 182, 543

\bibitem[\protect\citeauthoryear{Becker, White \& Helfand}{Becker
  et~al.}{1995}]{Beck95}
Becker R.~H.,  White R.~L.,    Helfand D.~J.,  1995, ApJ, 450, 559

\bibitem[\protect\citeauthoryear{{Best}}{{Best}}{2009}]{Best09}
{Best} P.~N.,  2009, Astronomische Nachrichten, 330, 184

\bibitem[\protect\citeauthoryear{Best, R\"ottgering \& Lehnert}{Best
  et~al.}{1999}]{Best99}
Best P.~N.,  R\"ottgering H. J.~A.,    Lehnert M.~D.,  1999, MNRAS, 310, 223

\bibitem[\protect\citeauthoryear{{Black}, {Baum}, {Leahy}, {Perley}, {Riley} \&
  {Scheuer}}{{Black} et~al.}{1992}]{Black92}
{Black} A.~R.~S.,  {Baum} S.~A.,  {Leahy} J.~P.,  {Perley} R.~A.,  {Riley}
  J.~M.,    {Scheuer} P.~A.~G.,  1992, \mnras, 256, 186

\bibitem[\protect\citeauthoryear{{Boroson} \& {Lauer}}{{Boroson} \&
  {Lauer}}{2009}]{Bor09}
{Boroson} T.~A.,  {Lauer} T.~R.,  2009, \nat, 458, 53

\bibitem[\protect\citeauthoryear{Brotherton}{Brotherton}{1996}]{Broth96}
Brotherton M.~S.,  1996, ApJS, 102, 1

\bibitem[\protect\citeauthoryear{Buttiglione, Capetti, Celotti, Axon,
  Chiaberge, Macchetto \& Sparks}{Buttiglione et~al.}{2009}]{But09}
Buttiglione S.,  Capetti A.,  Celotti A.,  Axon D.~J.,  Chiaberge M.,
  Macchetto F.~D.,    Sparks W.~B.,  2009, A\&A, 495, 1033

\bibitem[\protect\citeauthoryear{{Capetti}, {Zamfir}, {Rossi}, {Bodo}, {Zanni}
  \& {Massaglia}}{{Capetti} et~al.}{2002}]{Cap02}
{Capetti} A.,  {Zamfir} S.,  {Rossi} P.,  {Bodo} G.,  {Zanni} C.,
  {Massaglia} S.,  2002, \aap, 394, 39

\bibitem[\protect\citeauthoryear{Cardelli, Clayton \& Mathis}{Cardelli
  et~al.}{1989}]{Car89}
Cardelli J.~A.,  Clayton G.~C.,    Mathis J.~S.,  1989, ApJ, 345, 245

\bibitem[\protect\citeauthoryear{Cheung}{Cheung}{2007}]{Cheung07a}
Cheung C.~C.,  2007, AJ, 133, 2097

\bibitem[\protect\citeauthoryear{Cheung, Healey, Landt, Verdoes~Kleijn \&
  Jord\'an}{Cheung et~al.}{2009}]{Cheung09}
Cheung C.~C.,  Healey S.~E.,  Landt H.,  Verdoes~Kleijn G.,    Jord\'an A.,
  2009, ApJS, 181, 548

\bibitem[\protect\citeauthoryear{{Cheung} \& {Springmann}}{{Cheung} \&
  {Springmann}}{2007}]{Cheung07b}
{Cheung} C.~C.,  {Springmann} A.,  2007, in {L.~C.~Ho \& J.-W.~Wang} ed., The
  Central Engine of Active Galactic Nuclei Vol.~373 of Astronomical Society of
  the Pacific Conference Series, {FIRST ``Winged'' and ``X''-shaped Radio
  Source Candidates}.
p.~259

\bibitem[\protect\citeauthoryear{Corbin}{Corbin}{1997}]{Corb97}
Corbin M.~R.,  1997, ApJS, 113, 245

\bibitem[\protect\citeauthoryear{De~Robertis}{De~Robertis}{1985}]{DeR85}
De~Robertis M.,  1985, ApJ, 289, 67

\bibitem[\protect\citeauthoryear{Dennett-Thorpe, Scheuer, Laing, Bridle, Pooley
  \& Reich}{Dennett-Thorpe et~al.}{2002}]{Den02}
Dennett-Thorpe J.,  Scheuer P. A.~G.,  Laing R.~A.,  Bridle A.~H.,  Pooley
  G.~G.,    Reich W.,  2002, MNRAS, 330, 609

\bibitem[\protect\citeauthoryear{Dickey \& Lockman}{Dickey \&
  Lockman}{1990}]{DL90}
Dickey J.~M.,  Lockman F.~J.,  1990, ARA\&A, 28, 215

\bibitem[\protect\citeauthoryear{{Ekers}, {Fanti}, {Lari} \& {Parma}}{{Ekers}
  et~al.}{1978}]{Ekers78}
{Ekers} R.~D.,  {Fanti} R.,  {Lari} C.,    {Parma} P.,  1978, \nat, 276, 588

\bibitem[\protect\citeauthoryear{Ferland, Korista, Verner, Ferguson, Kingdon \&
  Verner}{Ferland et~al.}{1998}]{Cloudy}
Ferland G.~J.,  Korista K.~T.,  Verner D.~A.,  Ferguson J.~W.,  Kingdon J.~B.,
    Verner E.~M.,  1998, PASP, 110, 761

\bibitem[\protect\citeauthoryear{{Gaskell}}{{Gaskell}}{1996}]{Gas96}
{Gaskell} C.~M.,  1996, \apjl, 464, L107

\bibitem[\protect\citeauthoryear{{Gopal-Krishna}, {Biermann} \&
  {Wiita}}{{Gopal-Krishna} et~al.}{2003}]{Gop03}
{Gopal-Krishna} {Biermann} P.~L.,    {Wiita} P.~J.,  2003, \apjl, 594, L103

\bibitem[\protect\citeauthoryear{{Hodges-Kluck}, {Reynolds}, {Cheung} \&
  {Miller}}{{Hodges-Kluck} et~al.}{2010}]{Hodges10}
{Hodges-Kluck} E.~J.,  {Reynolds} C.~S.,  {Cheung} C.~C.,    {Miller} M.~C.,
  2010, \apj, 710, 1205

\bibitem[\protect\citeauthoryear{{Kraft}, {Hardcastle}, {Worrall} \&
  {Murray}}{{Kraft} et~al.}{2005}]{Kraft05}
{Kraft} R.~P.,  {Hardcastle} M.~J.,  {Worrall} D.~M.,    {Murray} S.~S.,  2005,
  \apj, 622, 149

\bibitem[\protect\citeauthoryear{Lacy}{Lacy}{2000}]{Lacy00}
Lacy M.,  2000, ApJ, 536, L1

\bibitem[\protect\citeauthoryear{Laing, Jenkins, Wall \& Unger}{Laing
  et~al.}{1994}]{Lai94}
Laing R.~A.,  Jenkins C.~R.,  Wall J.~V.,    Unger S.~W.,  1994, in Bicknell
  G.~V.,  Dopita M.~A.,   Quinn P.~J.,  eds, The First Stromlo Symposium: The
  Physics of Active Glaxies Spectrophotometry of a complete sample of {3CR}
  radio sources: Implications for unified models.
A.S.P., San Francisco, p.~201

\bibitem[\protect\citeauthoryear{{Lal} \& {Rao}}{{Lal} \& {Rao}}{2007}]{Lal07}
{Lal} D.~V.,  {Rao} A.~P.,  2007, \mnras, 374, 1085

\bibitem[\protect\citeauthoryear{Landt \& Bignall}{Landt \&
  Bignall}{2008}]{L08c}
Landt H.,  Bignall H.~E.,  2008, MNRAS, 391, 967

\bibitem[\protect\citeauthoryear{Landt, Padovani \& Giommi}{Landt
  et~al.}{2002}]{L02}
Landt H.,  Padovani P.,    Giommi P.,  2002, MNRAS, 336, 945

\bibitem[\protect\citeauthoryear{Landt, Padovani, Perlman \& Giommi}{Landt
  et~al.}{2004}]{L04}
Landt H.,  Padovani P.,  Perlman E.~S.,    Giommi P.,  2004, MNRAS, 351, 83

\bibitem[\protect\citeauthoryear{Leahy \& Parma}{Leahy \&
  Parma}{1992}]{Leahy92}
Leahy J.~P.,  Parma P.,  1992, in Roland J.,  Sol H.,   Pelletier G.,  eds,
  Extragalactic Radio Sources. From Beams to Jets Multiple outbursts in radio
  galaxies.
Cambridge University Press, p.~307

\bibitem[\protect\citeauthoryear{{Leahy} \& {Williams}}{{Leahy} \&
  {Williams}}{1984}]{Leahy84}
{Leahy} J.~P.,  {Williams} A.~G.,  1984, \mnras, 210, 929

\bibitem[\protect\citeauthoryear{Ledlow \& Owen}{Ledlow \& Owen}{1996}]{Led96}
Ledlow M.~J.,  Owen F.~N.,  1996, AJ, 112, 9

\bibitem[\protect\citeauthoryear{Mannucci, Basile, Poggianti, Cimatti, Daddi,
  Pozzetti \& Vanzi}{Mannucci et~al.}{2001}]{Man01}
Mannucci F.,  Basile F.,  Poggianti B.~M.,  Cimatti A.,  Daddi E.,  Pozzetti
  L.,    Vanzi L.,  2001, MNRAS, 326, 745

\bibitem[\protect\citeauthoryear{Marziani, Sulentic, Dultzin-Hacyan, \~Calvani
  \& Moles}{Marziani et~al.}{1996}]{Marz96}
Marziani P.,  Sulentic J.~W.,  Dultzin-Hacyan D.,  \~Calvani M.,    Moles M.,
  1996, ApJS, 104, 37

\bibitem[\protect\citeauthoryear{{Merritt} \& {Ekers}}{{Merritt} \&
  {Ekers}}{2002}]{Mer02}
{Merritt} D.,  {Ekers} R.~D.,  2002, Science, 297, 1310

\bibitem[\protect\citeauthoryear{{Miller} \& {Brandt}}{{Miller} \&
  {Brandt}}{2009}]{Miller09}
{Miller} B.~P.,  {Brandt} W.~N.,  2009, \apj, 695, 755

\bibitem[\protect\citeauthoryear{{Murgia}, {Parma}, {de Ruiter}, {Bondi},
  {Ekers}, {Fanti} \& {Fomalont}}{{Murgia} et~al.}{2001}]{Murgia01}
{Murgia} M.,  {Parma} P.,  {de Ruiter} H.~R.,  {Bondi} M.,  {Ekers} R.~D.,
  {Fanti} R.,    {Fomalont} E.~B.,  2001, \aap, 380, 102

\bibitem[\protect\citeauthoryear{Osterbrock \& Ferland}{Osterbrock \&
  Ferland}{2006}]{Osterbrock2}
Osterbrock D.~E.,  Ferland G.~J.,  2006, Astrophysics of Gaseous Nebulae and
  Active Galactic Nuclei.
University Science Books

\bibitem[\protect\citeauthoryear{Perlman, Padovani, Giommi, Sambruna, Jones,
  Tzioumis \& Reynolds}{Perlman et~al.}{1998}]{Per98}
Perlman E.~S.,  Padovani P.,  Giommi P.,  Sambruna R.,  Jones L.~R.,  Tzioumis
  A.,    Reynolds J.,  1998, AJ, 115, 1253

\bibitem[\protect\citeauthoryear{Peterson}{Peterson}{1997}]{Peterson}
Peterson B.~M.,  1997, An Introduction to Active Galactic Nuclei.
Cambridge University Press

\bibitem[\protect\citeauthoryear{{Peterson}, {Korista} \& {Cota}}{{Peterson}
  et~al.}{1987}]{Pet87}
{Peterson} B.~M.,  {Korista} K.~T.,    {Cota} S.~A.,  1987, \apjl, 312, L1

\bibitem[\protect\citeauthoryear{Rawlings \& Saunders}{Rawlings \&
  Saunders}{1991}]{Raw91}
Rawlings S.,  Saunders R.,  1991, Nature, 349, 138

\bibitem[\protect\citeauthoryear{Rottmann}{Rottmann}{2001}]{Rott01}
Rottmann H.,  2001, PhD thesis, Rheinische Friedrich-Wilhelms-Universit\"at
  Bonn

\bibitem[\protect\citeauthoryear{{Saripalli} \& {Subrahmanyan}}{{Saripalli} \&
  {Subrahmanyan}}{2009}]{Sar09}
{Saripalli} L.,  {Subrahmanyan} R.,  2009, \apj, 695, 156

\bibitem[\protect\citeauthoryear{{Shen} \& {Loeb}}{{Shen} \&
  {Loeb}}{2009}]{Shen09}
{Shen} Y.,  {Loeb} A.,  2009, ArXiv e-prints

\bibitem[\protect\citeauthoryear{Tadhunter, Morganti, di Serego-Alighieri,
  Fosbury \& Danziger}{Tadhunter et~al.}{1993}]{Tad93}
Tadhunter C.~N.,  Morganti R.,  di Serego-Alighieri S.,  Fosbury R. A.~E.,
  Danziger I.~J.,  1993, MNRAS, 263, 999

\bibitem[\protect\citeauthoryear{Tadhunter, Morganti, Robinson, Dickson,
  Villar-Martin \& Fosbury}{Tadhunter et~al.}{1998}]{Tad98}
Tadhunter C.~N.,  Morganti R.,  Robinson A.,  Dickson R.,  Villar-Martin M.,
  Fosbury R. A.~E.,  1998, MNRAS, 298, 1035

\bibitem[\protect\citeauthoryear{{Ulrich} \& {Roennback}}{{Ulrich} \&
  {Roennback}}{1996}]{Ulr96}
{Ulrich} M.,  {Roennback} J.,  1996, \aap, 313, 750

\bibitem[\protect\citeauthoryear{{van Breugel}, {Balick}, {Heckman}, {Miley} \&
  {Helfand}}{{van Breugel} et~al.}{1983}]{Breu83}
{van Breugel} W.,  {Balick} B.,  {Heckman} T.,  {Miley} G.,    {Helfand} D.,
  1983, \aj, 88, 40

\bibitem[\protect\citeauthoryear{{Wang}, {Zhou} \& {Dong}}{{Wang}
  et~al.}{2003}]{Wang03}
{Wang} T.,  {Zhou} H.,    {Dong} X.,  2003, \aj, 126, 113

\bibitem[\protect\citeauthoryear{{Wirth}, {Smarr} \& {Gallagher}}{{Wirth}
  et~al.}{1982}]{Wirth82}
{Wirth} A.,  {Smarr} L.,    {Gallagher} J.~S.,  1982, \aj, 87, 602

\bibitem[\protect\citeauthoryear{{Worrall}, {Birkinshaw} \&
  {Cameron}}{{Worrall} et~al.}{1995}]{Wor95}
{Worrall} D.~M.,  {Birkinshaw} M.,    {Cameron} R.~A.,  1995, \apj, 449, 93

\bibitem[\protect\citeauthoryear{{Zhang}, {Dultzin-Hacyan} \& {Wang}}{{Zhang}
  et~al.}{2007}]{Zhang07}
{Zhang} X.,  {Dultzin-Hacyan} D.,    {Wang} T.,  2007, \mnras, 377, 1215

\end{thebibliography}

\appendix

\section{SDSS spectra}


\begin{figure*}
\centerline{
\includegraphics[scale=1.0]{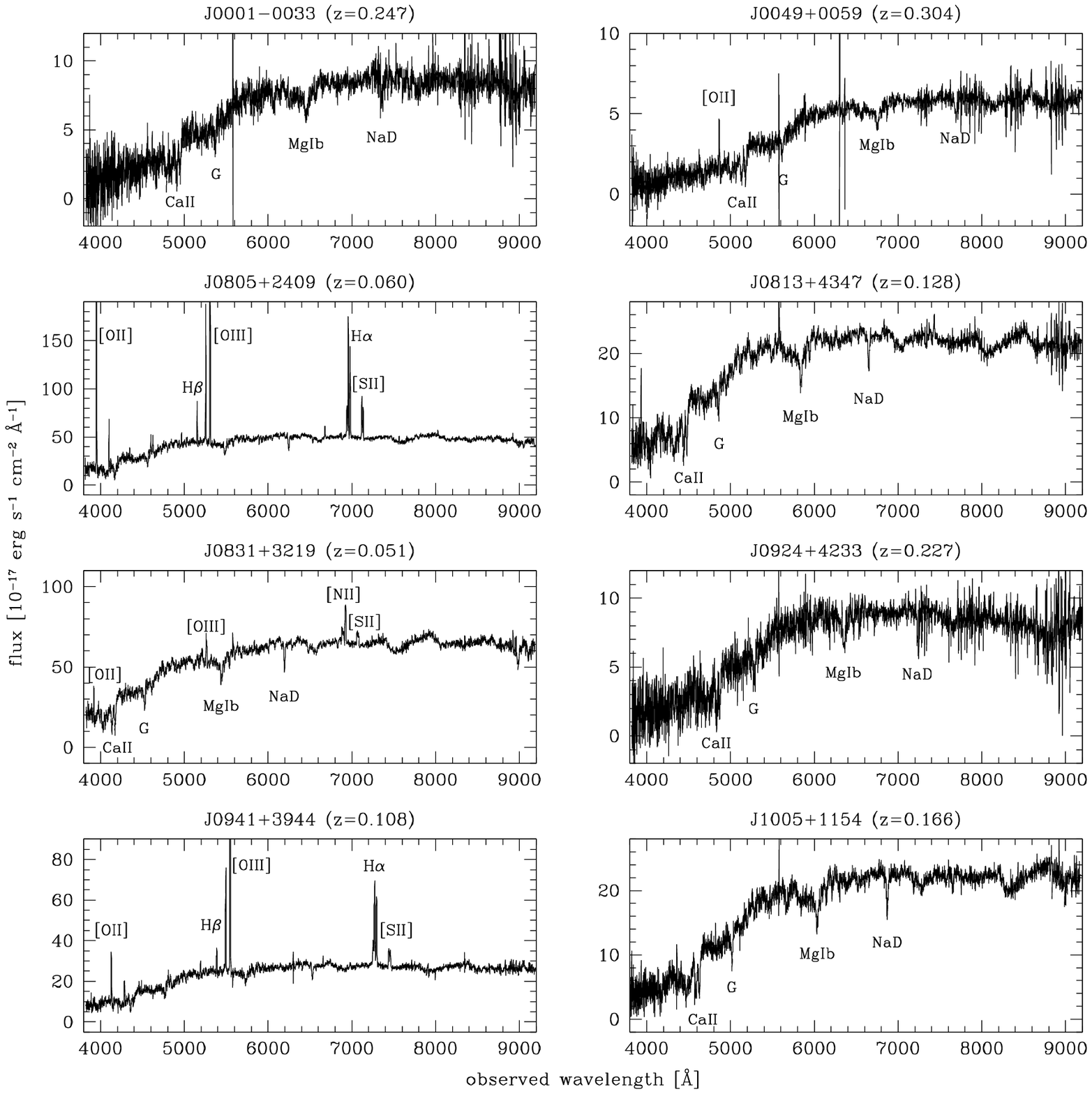}
}
\caption{\label{sdss} Observed Sloan Digital Sky Survey (SDSS) spectra
(corrected for Galactic extinction) of X-shaped radio sources.}
\end{figure*}

\setcounter{figure}{1}
\begin{figure*}
\centerline{
\includegraphics[scale=1.0]{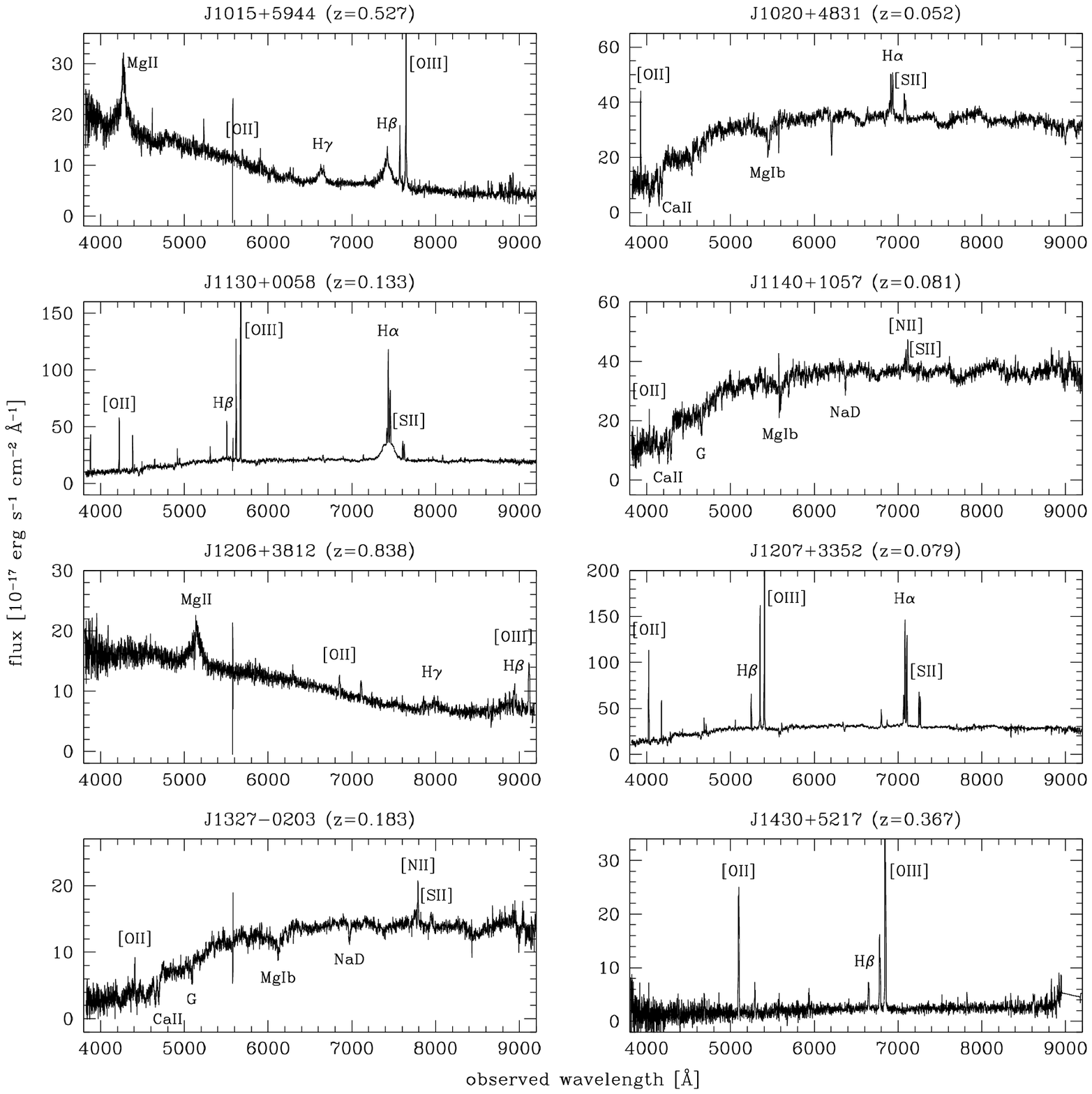}
}
\contcaption{}
\end{figure*}

\setcounter{figure}{1}
\begin{figure*}
\centerline{
\includegraphics[scale=1.0]{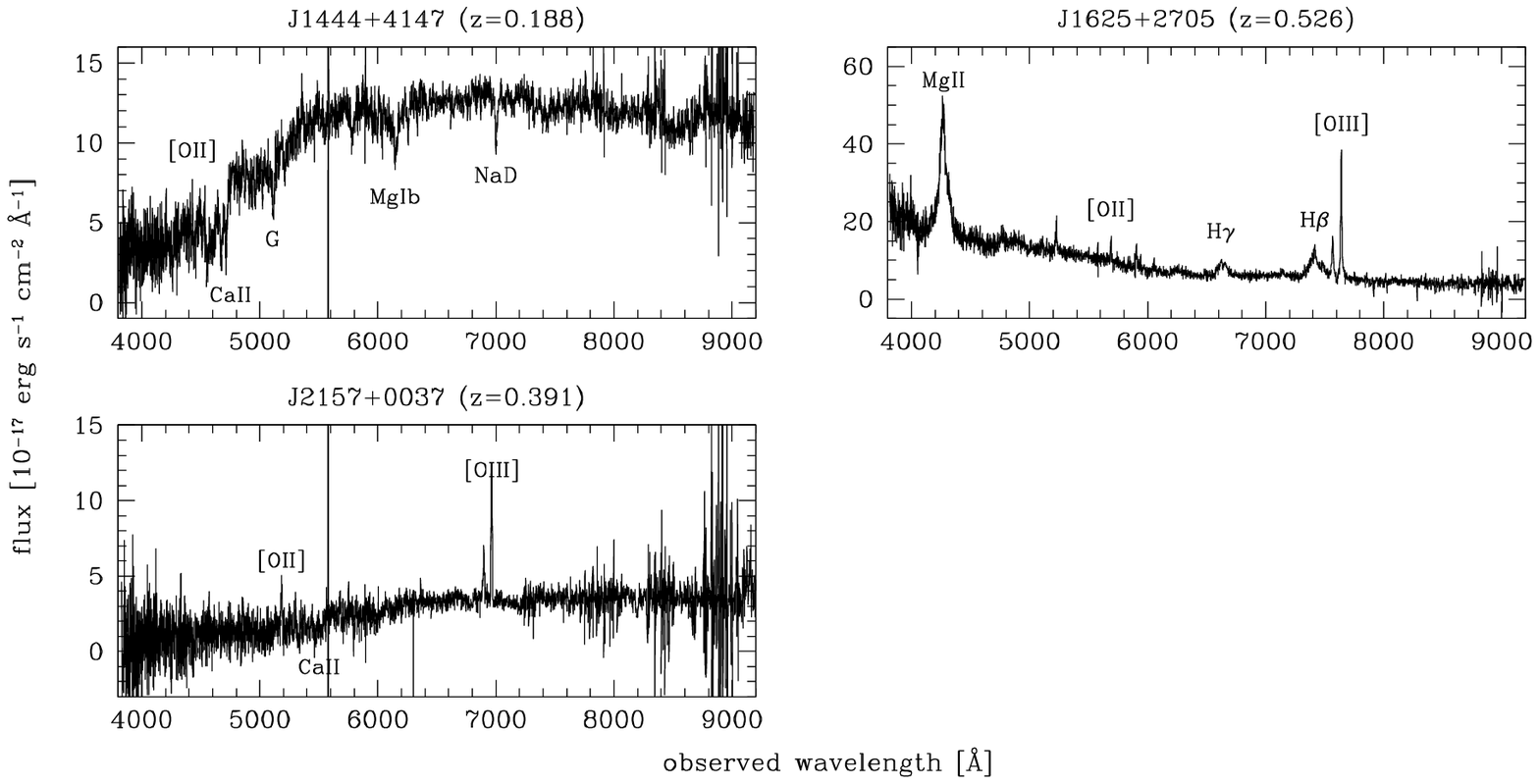}
}
\contcaption{}
\end{figure*}

\bsp
\label{lastpage}

\end{document}